\documentclass{aa}

\usepackage{graphicx}
\usepackage{txfonts}

\usepackage{academicons}

\usepackage{natbib,twoopt}
\usepackage[dvipsnames]{xcolor}

\newcommand{\Gaia}{{\it Gaia}}
\newcommand{\flags}{{\it flags\_gspspec}}
\newcommand{\gspspec}{{\it GSP-Spec}}

\newcommand{\GamDor}{$\gamma$~Dor}

\newcommand{\T}{$T_{\rm eff}$}
\newcommand{\g}{log($g$)}

\newcommand{\meta}{[M/H]}
\newcommand{\alfa}{$\alpha$}
\newcommand{\AF}{[\alfa/Fe]}

\newcommand{\Vrad}{$V_{\rm Rad}$}

\newcommand{\SNR}{$S/N$}

\newcommand{\CaII}{Ca~{\sc ii}}

\newcommand{\CaFe}{[Ca/Fe]}

\newcommand{\FeH}{[Fe/H]}

\usepackage{listings}
\definecolor{dkgreen}{rgb}{0,0.6,0}
\definecolor{gray}{rgb}{0.5,0.5,0.5}
\definecolor{mauve}{rgb}{0.58,0,0.82}

\begin{document} 

   \title{\Gaia/GSP-spec spectroscopic properties of $\gamma$~Doradus 
pulsators}

   \author{P. de Laverny \inst{1}
          \and
          A. Recio-Blanco\inst{1}
          \and
          C. Aerts \inst{2,3,4}
          \and
          P.A. Palicio\inst{1}
          } 

   \institute{
   Université Côte d'Azur, Observatoire de la Côte d'Azur, CNRS, Laboratoire Lagrange, Bd de l'Observatoire, CS 34229, 06304 Nice cedex 4, France  
   \and
Institute of Astronomy, KU Leuven, Celestijnenlaan 200D, 3001, Leuven, Belgium
      \and
      Department of Astrophysics, IMAPP, Radboud University Nijmegen, PO Box 9010, 6500 GL Nijmegen, The Netherlands
      \and
      Max Planck Institut für Astronomie, Königstuhl 17, 69117 Heidelberg, Germany
   }

   \date{Received ?? ; accepted ??}

   \abstract
   {The third Data Release of the ESA \Gaia\ mission has provided 
   a large sample of new gravity-mode pulsators, among which more than 11,600 are \GamDor\ stars.}
   {The goal of the present work is to present  the spectroscopic parameters of these \GamDor\ pulsators estimated by the the \gspspec\ module that analysed millions of \Gaia\ spectra. Such a parametrisation could help to confirm their \GamDor\ nature and provide their chemo-physical properties.}
   {The Galactic positions, kinematics, and orbital properties of these new \Gaia\ pulsators were examined in order to define a sub-sample belonging to the Milky Way thin disc, in which these young stars should preferentially be found. The stellar luminosities, radii, and astrometric surface gravities were estimated without adopting any priors from uncertain stellar evolution models. These parameters, combined with the \gspspec\ effective temperatures, spectroscopic gravities, and metallicities were then validated by comparison with recent literature studies.}
   {Most stars are found to belong to the Galactic thin disc, as expected. It is also found that the derived luminosities, radii, and astrometric surface gravities are of high quality and have values typical of genuine \GamDor\ pulsators. Moreover, we have shown that \T\ and \meta\ of pulsators with high enough \SNR\ spectra or slow to moderate rotation rates are robust. This allowed to define a sub-sample of genuine slow-rotating  \Gaia\ \GamDor\ pulsators. Their \T\ are found between $\sim$6,500 and  
$\sim$7,800~K, \g\ around 4.2 and luminosities and stellar radii peak at $\sim$5~L$_\odot$ and $\sim$1.7~L$_\odot$, The median metallicity is close to the Solar value, although 0.5~dex more metal-poor and metal-rich \GamDor\ are identified.
   The \AF\ content  is fully consistent with the chemical properties of the Galactic  disc.}
   {\Gaia/DR3 spectroscopic properties of \GamDor\ stars therefore confirm the nature of these pulsators and allow to chemo-physically parametrise a new large sample of such stars. 
   Moreover, future \Gaia\ data releases should drastically increase the number of \GamDor\ stars with 
   good-precision spectroscopically derived parameters.}
   \keywords{Asteroseismology – Stars: rotation, oscillations (including pulsation), fundamental parameters, abundances - Galaxy: abundances, surveys}   
   \maketitle
   
\section{Introduction}
\label{Sec:Intro}

The European Space Agency \Gaia\ mission with the release of its third catalogue \citep[DR3,][]{DR3Antonella} has already revolutionised different fields of astrophysics. In particular our view of the Milky Way stellar populations
is being upgraded substantially \citep{PVP_Ale}.
Regarding stellar physics and variable stars, 
\cite{Conny23} assessed the  fundamental parameters and mode properties of 15,602 newly found \Gaia/DR3 gravity-mode 
($g$-mode hereafter)
pulsator candidates, among the more than 100,000 new pulsators along the main sequence identified from their \Gaia\ photometric light-curves 
by \citet{DeRidder23}.
These $g$-mode pulsators are low- and intermediate-mass main-sequence stars (masses between 
1.3 and 9~M$_\sun$) and are ideal laboratories for 
asteroseismology, within the broad landscape of such modern studies
\citep[see][for recent reviews]{Aerts21,Kurtz22}.

Meanwhile \citet{HeyAerts24} extracted light curves assembled with the Transiting Exoplanet Survey Satellite \citep[TESS,][]{Ricker15} for more than 60,000 of the candidate pulsators discovered by \citet{DeRidder23}. They confirmed the pulsational nature for the large majority and found them to be multiperiodic, with about 70\% of them even sharing the same dominant frequency in the totally independent \Gaia\ DR3 and TESS data. By comparing the astrophysical 
properties of the \Gaia\ $g$-mode pulsators with those studied 
asteroseismically from {\it Kepler\/} data by 
\citet[][for the B-type stars]{Pedersen21} and by 
\citet[][for the F-type stars]{VanReeth16,GangLi20}, \citet{Conny23} classified them either as 
Slowly Pulsating B (SPB) or $\gamma$~Doradus (\GamDor) candidate stars. 
Our current work is focused on this last category of pulsating F-type dwarfs, covering masses
$1.3\,$M$_\odot\la$ M $\la 1.9$~M$_\odot$. In the Hertzsprung-Russell diagram (HRD), they are
found in a rather small main-sequence area
\citep[][Fig.\,2]{Fritzewski24},
close to the cool limit of $\delta$~Scuti instability strip \citep{Murphy19}.
However, we do point out that many of the \GamDor\ stars actually turn out to be hybrid pulsators when observed in high-cadence space photometry. Indeed a good fraction of these pulsators exhibit not only high-order $g$~modes, but also acoustic waves known as  pressure (or $p$)~modes \citep{Achmed10,Hareter11,Arias17,Audenaert22}. This makes their instability strip overlap with the one of the $\delta$~Scuti stars \citep{DeRidder23,HeyAerts24}.

One of the new products published with \Gaia/DR3 are the stellar atmospheric parameters 
derived from the analysis of the \Gaia/Radial Velocity Spectrometer (RVS) spectra by the DPAC/\gspspec\ module \citep{GSPspecDR3}. 
 RVS spectra cover the Ca~II IR domain   (846–870~nm) and have a resolution around 11,500.
By automatically analysing these spectra, \cite{GSPspecDR3} parametrised about 5.6 million single low-rotating stars
belonging to the FGKM-spectral type. Hotter stars (\T>8000~K)  or highly-rotating stars (more than $\sim$30-50~km.s$^{-1}$, depending on the stellar type) were disregarded for the \Gaia/DR3 (this will be updated for the next \Gaia\ Data Releases). These limitations come from the adopted reference grids upon which rely the parametrisation algorithms of \gspspec.
The derived stellar atmospheric parameters are: the effective temperature \T, the surface gravity \g, the global metallicity \meta, and the enrichment in $\alpha$-element with respect to iron \AF. Moreover, up to 13 individual chemical abundances were also estimated for most of these stars. This analysis led to the first all-sky spectroscopic catalogue and the largest compilation of stellar chemo-physical parameters ever published. Moreover, radial velocities (\Vrad) of about 33 millions stars were published in the DR3 catalogue \citep{Katz23}. All these data allow to study the Galactic kinematics and orbital properties 
of this huge number of stars, along
with their atmospheric and chemical characteristics, 
keeping in mind that they belong
to various populations in the Milky Way. This multitude of data also facilitates constraining stellar evolution models for 
a broad range of masses. In particular, the \Gaia\ spectroscopic data give us the opportunity to unravel the properties of different kinds of variable stars, among which the $g$-mode pulsators
characterised photometrically by \citet{Conny23}.

One of the main goals of the present study is to apply 
spectroscopic techniques to characterise the $g$-mode 
and hybrid
pulsators identified in \cite{Conny23}
and \citet{HeyAerts24}.
This perspective adds to their photometrically-deduced properties and may help future asteroseismic modelling of the most promising pulsators.
Some contamination by other or additional 
variability could also occur, such as 
rotational modulation.  This is expected since 
\citet{DeRidder23} and \citet{Conny23} 
could infer only one secure frequency  for many of these 15,602 candidates. Moreover, \citet{HeyAerts24} found
a fraction of the $g$-mode 
and hybrid
pulsators to reveal Ap/Bp characteristics in addition to their pulsational behaviour. 
Therefore, some of them might be Ap/Bp pulsators with anomalous chemical abundances from spots, as previously found from {\it Kepler\/} space photometry \citep[e.g.,][]{Bowman18,Henriksen23}.
\Gaia/\gspspec\ spectroscopic parameters could therefore help to confirm the nature of these pulsators and facilitate to build sub-catalogues of $g$-mode 
and hybrid
pulsating stars with respect to their chemical properties.
In addition, the main properties of these stars can be constrained from the \Gaia\ spectroscopy and compared with the values deduced from the photometry, such as their effective temperature, surface gravity, luminosities, radius, etc.

Another goal is to explore the metallicity (and, possibly, chemical abundances)  of genuine \Gaia\  $g$-mode 
and hybrid
pulsators. This is an important input for asteroseismic modelling, once a sufficient number of oscillation frequencies has been identified\citep{Aerts18,Mombarg21,Mombarg22}.
Aside from the {\it Kepler\/} sample of \GamDor\ pulsators modelled by \citet{GangLi20}, the confirmed $g$-mode 
and hybrid
pulsators from \citet{HeyAerts24} are currently  under study 
to derive their global parameters
(mass, convective core mass, radius, and evolutionary stage; 
Mombarg et al., submitted), as well as 
all their significant oscillation mode frequencies from high-cadence 
TESS photometry. Future studies will point  out whether their TESS data allow us to find period spacing patterns to assess their suitability for asteroseismology, as has been possible for \GamDor\ stars in the TESS Continuous Viewing Zones \citep{Garcia22a,Garcia22b}.

We note that the present work focuses only on the \GamDor\ stars 
or hybrid pulsators with dominant $g$~modes
in the catalogue by \citet{Conny23}, for several reasons.
First, only $\sim$12\% of the SPB candidates 
in that paper have a DR3 radial
velocity (\Vrad) and, when available, the associated \Vrad\ uncertainties are quite large,
suggesting possible problems in their RVS spectra.
Moreover, most of the SPB candidates
are too faint to have a high enough \SNR\ to be analysed meaningfully with
\gspspec. Finally, we remind that \gspspec\ 
was initially constructed to achieve proper parametrisation of slowly-rotating rather cool stars. In particular,
the reference training grid does not contain model spectra for stars
hotter than \T$\sim$8000~K and the lines identified in the stellar spectra are assumed not to be too broadened. These restrictions led to the rejection of A-type or hotter stars and/or to lower quality parametrisation for the 
fastest
rotators, as indicated by specific quality flags (and even rejection for the extreme cases). This is confirmed by examining the few SPB 
stars with available \gspspec\ parameters: most of them are indeed flagged as doubtful. We therefore postpone the spectroscopic study of these SPB to the next \Gaia\ Data Releases in which the spectral parametrisation of hotter and/or fast rotators, such as many of the \GamDor\ and SPB pulsators \citep[see ][for summaries]{Aerts19,GangLi20,Pedersen21,Aerts21},
will be optimised.

This article is structured as follows.
In Sect.~\ref{Sec:GaiaSample}, we present the sample of \GamDor\ pulsators with available \Gaia\ spectroscopic data and we study their spatial distribution, kinematics, and orbital properties in the Milky Way. This led to the definition of a sub-sample of stars with very-high probability of being genuine \GamDor\ stars belonging to the Galactic thin disc.
Subsequently, we present in Sect.~\ref{Sec:Param} the physical parameters of the \GamDor\ candidates analysed by the \gspspec\ module and discuss their properties in Sect.~\ref{Sec:properties}.
Our conclusions are summarised in Sect.~\ref{Conclu}.

\begin{figure*}[ht!]
        \centering
        \includegraphics[scale = 0.22]{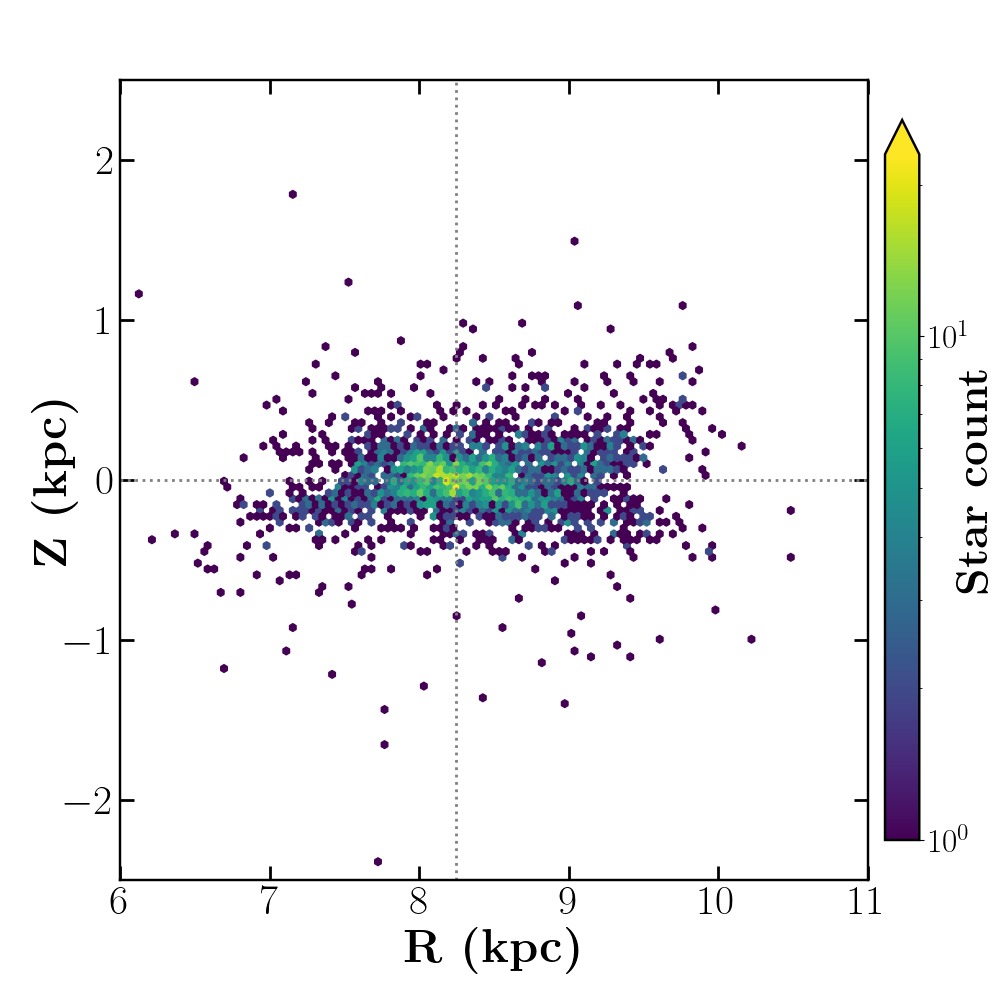}
        \includegraphics[scale = 0.22]{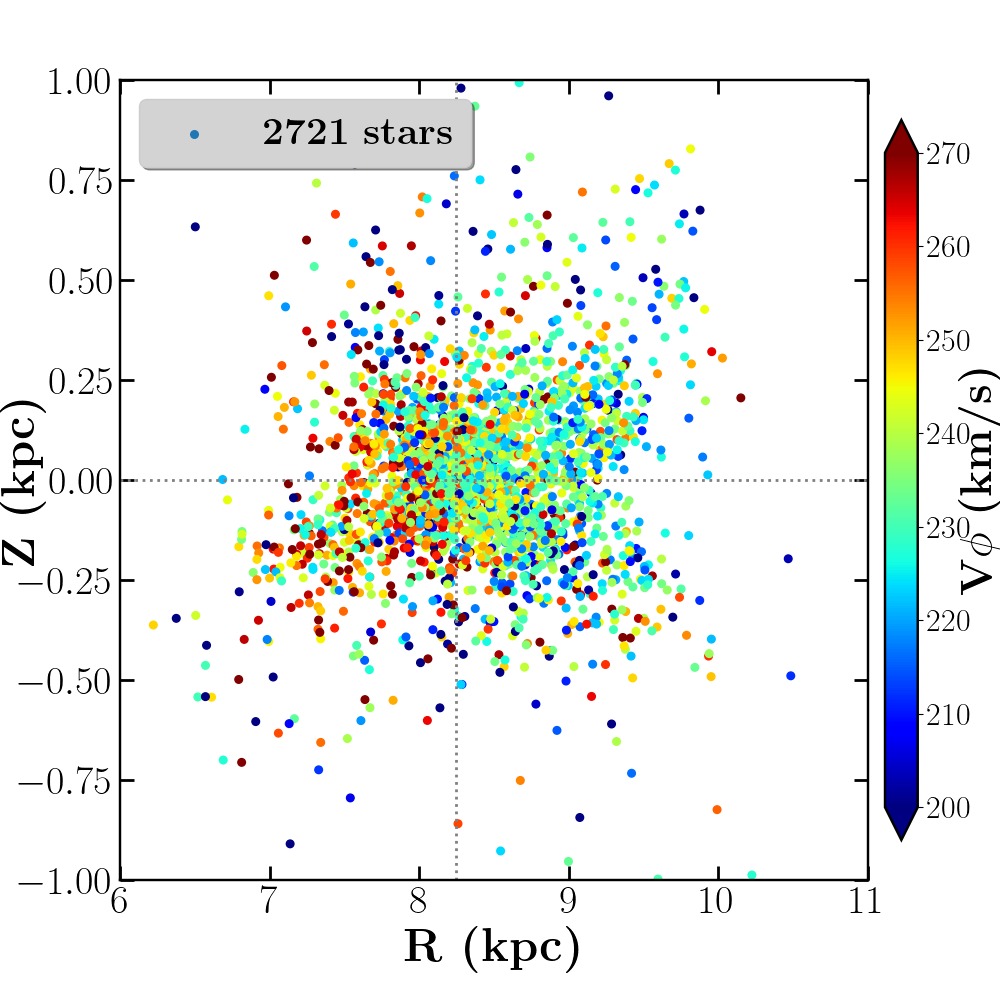}
        \includegraphics[scale = 0.22]{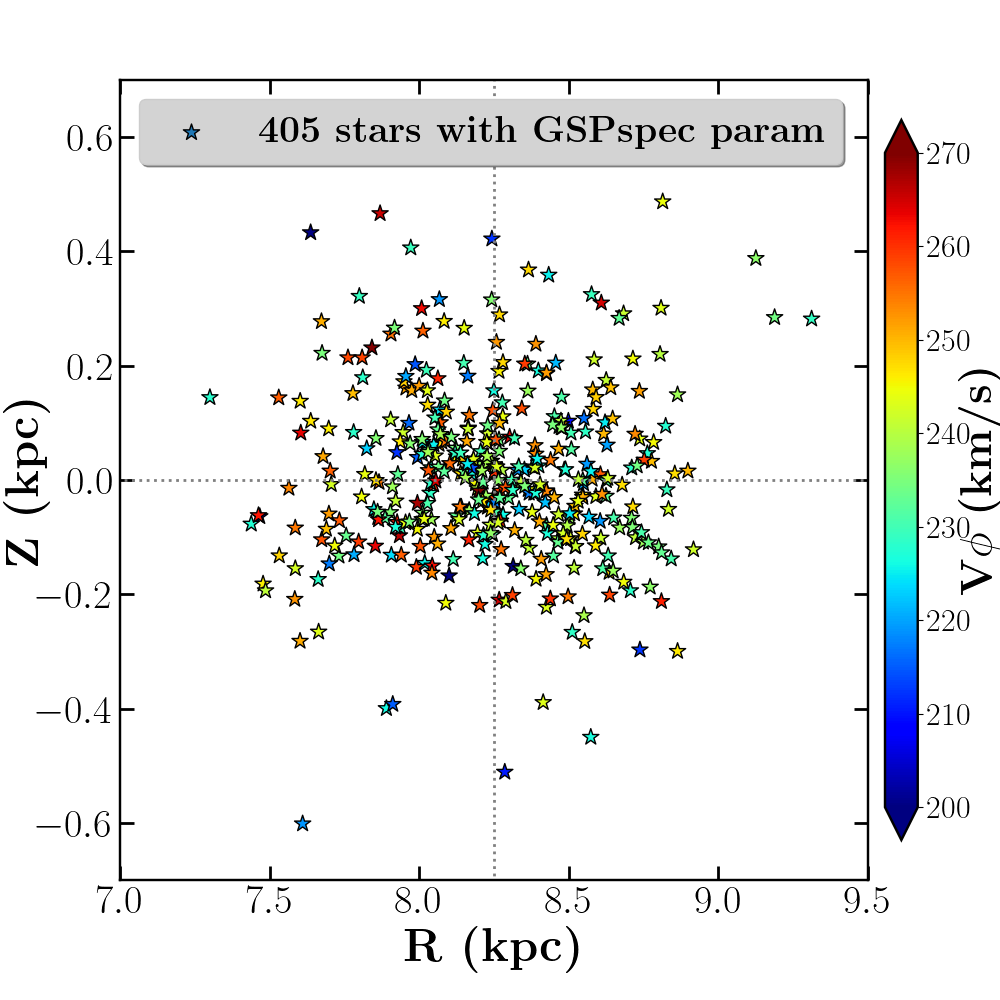}  
        \caption{Galactic location of the 2,721 \GamDor\ candidates with high-quality astrometric and radial velocity data. The Solar position is indicated by the dotted lines. The left panel is a density plot of the whole sample and the middle and right panels colour-code shows their Galactic rotational velocity ($V_\phi$). The right panel illustrates the location of the stars parametrised by \gspspec. The domain of shown (R, Z)-values in each panel decreases from left to right.}
        \label{Fig:RZ}
\end{figure*}

\begin{figure}[h!]
        \centering
        \includegraphics[scale = 0.25]{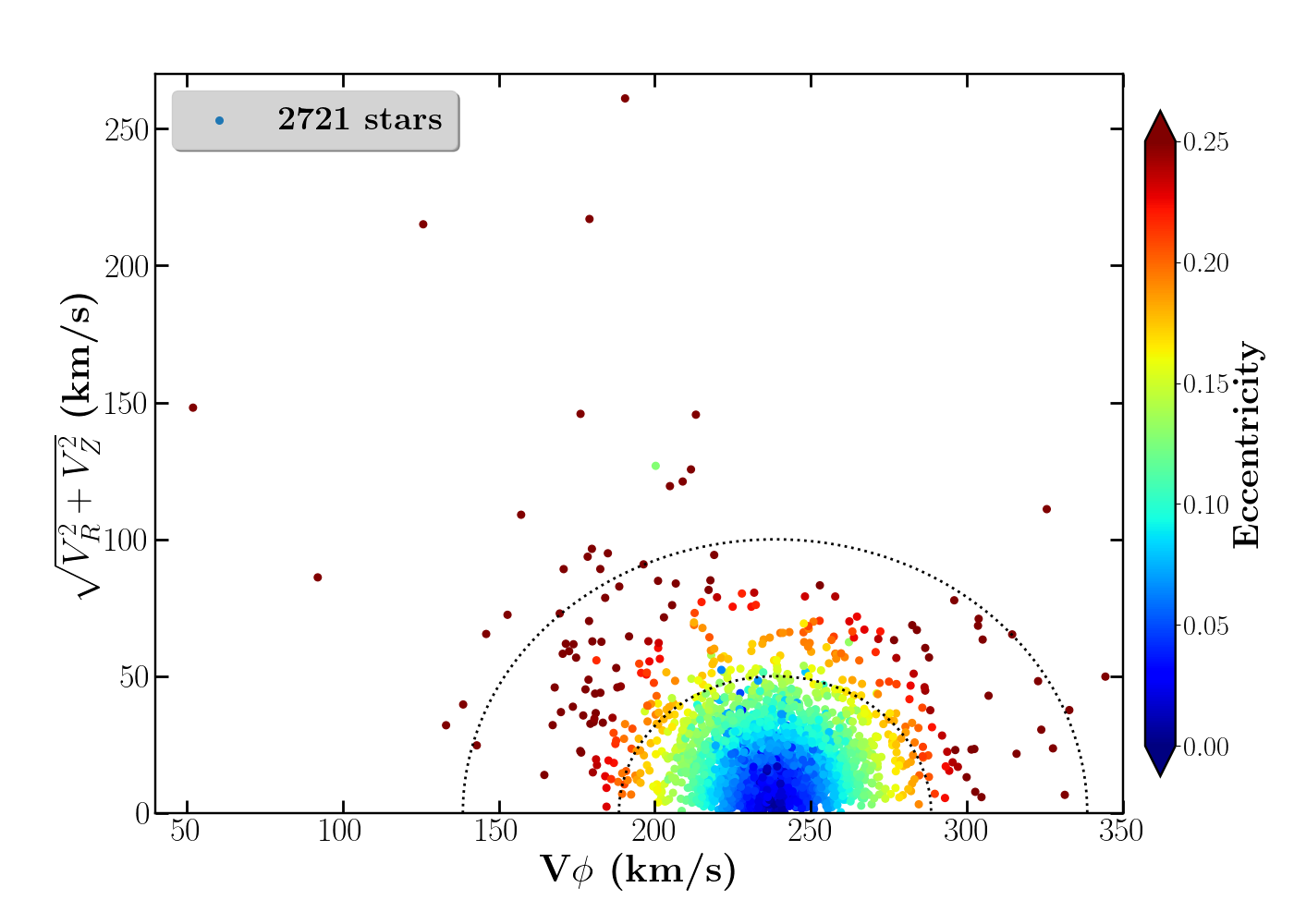}
        \includegraphics[scale = 0.25]{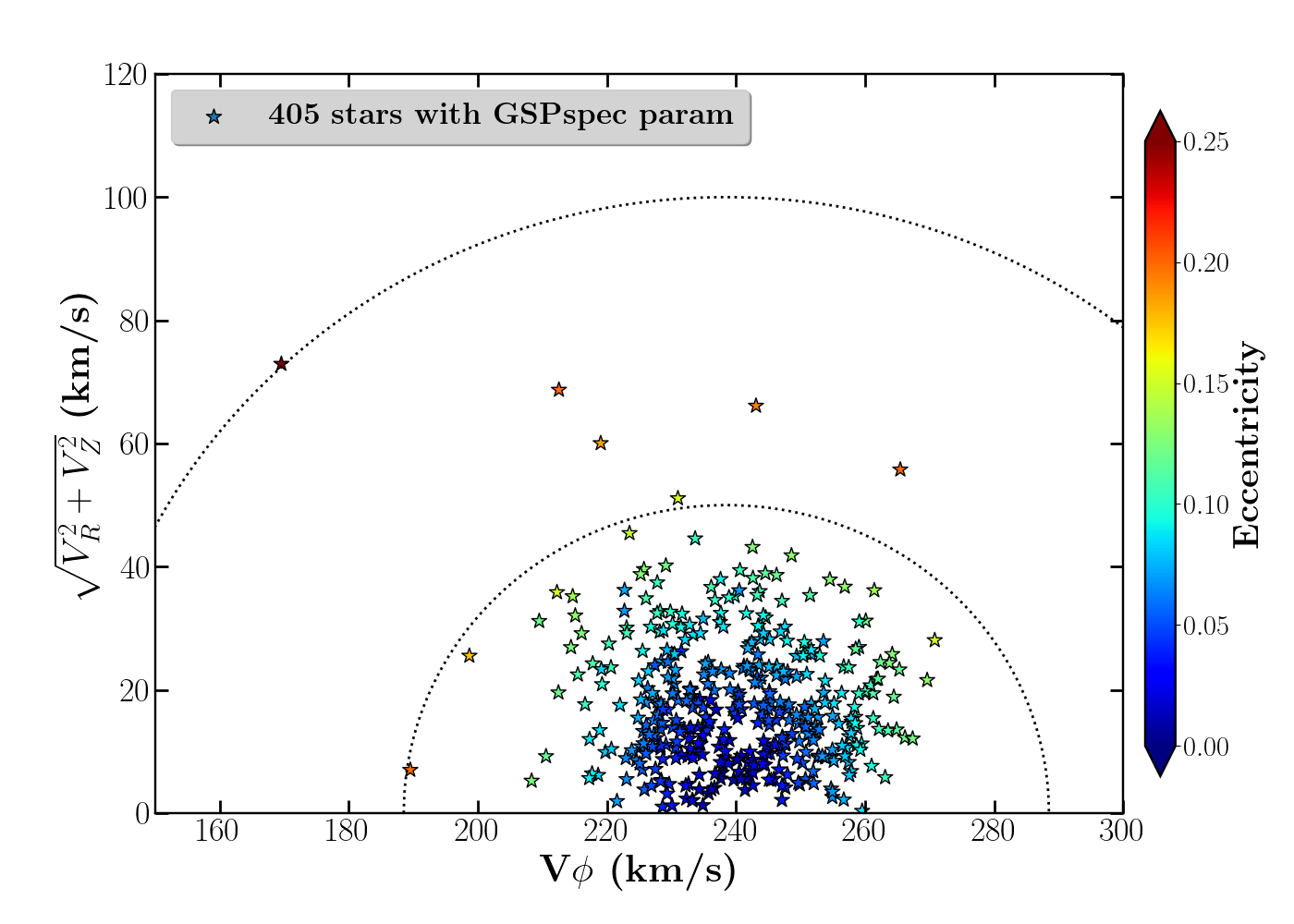}
        \caption{Toomre diagrams of the 2,721 \GamDor\ candidates with high-quality astrometric and radial velocity data, colour-coded with their orbital eccentricity. The upper and lower panels show the whole sample and a closer view of the sub-sample of stars parametrised by \gspspec, respectively. The dotted lines correspond to a total velocity equal to 50 and 100~km/s with respect to the LSR value. }
        \label{Fig:Toomre}
\end{figure}

\section{The \Gaia/DR3 spectroscopic sample of \GamDor\ stars}
\label{Sec:GaiaSample}

Among the 11,636 \GamDor\ pulsator candidates presented in \cite{Conny23}, 4,383 stars (38\%) have a published \Gaia/DR3 radial velocity \citep{Katz23} and only 650 of them are found in the \gspspec\ catalogue. 
These numbers can be explained by taking into account that (i) 58\% of these pulsating stars are too faint for having a high enough \SNR\ spectrum,
necessary to estimate their \Vrad\ and/or to be parametrised by \gspspec\ ($G\la$13.5~mag for this later case) and (ii) many of them are too hot,  preventing their \gspspec\ parametrisation (see the limitation caused by the reference grid hot boundary, described in the introduction).
In the following, we will discuss the Galactic properties of these candidate stars, derived from their available distances and \Vrad, in order to 
define a sub-sample of bona-fide \GamDor\ stars based only on kinematic and dynamical criteria.

\subsection{Spatial distribution, kinematics, and Galactic orbits}
\label{Sec:ThinDisc}
We derived the spatial (Cartesian coordinates) and kinematic properties of all the 4,383 \GamDor\ candidate stars from their \Gaia\ coordinates, proper motions and \Vrad,  adopting the 
distances of \cite{Coryn21}. Their Galactic orbital properties (eccentricity) were computed as described in \cite{Pedro23} using the Solar Galactic constants presented in \cite{PVP_Ale}. 
We also adopted for the LSR velocity of the Sun $V_{\rm LSR}$=238.5~km/s.

Among all these stars with Galactic data, 560 were parametrised by \gspspec. We note that the parametrisation is available for 90 more stars (see Sect.~\ref{Sec:Param}) because \gspspec\ parametrised some spectra whose \Vrad\ was finally not published within \Gaia/DR3, hence their Galactic properties were not computed.

The following quality selections were then applied to define a sub-sample of 2,721 stars (405 of them with \gspspec\ parameters) with high-quality Galactic parameters. (1) The best astrometric data were selected thanks to the $ruwe$ parameter ($ruwe$ < 1.4) and the identification of the non-spurious
solutions \citep[$fidelity_{v2}$ > 0.5, ][]{Rybizki22}.
This filters out 521 stars.
(2) We then rejected 152 stars having a distance uncertainty larger than 10\%.
(3) The \Vrad\ determination of several of these \GamDor\ candidates was found to be of poor quality, mostly because of the low \SNR\ of their spectra. We therefore disregarded 1,258 stars having a relative \Vrad\ error larger than 50\%. 

The Galactic location of this high-quality sub-sample of 2,721 \GamDor\ candidates with astrometric and \Vrad\ information is presented in Fig.\ref{Fig:RZ}. The colour-code of the middle and right panels represents their rotational velocity in the Galactic plane ($V_\phi$, whose typical value for thin disc stars in the Solar vicinity is around $\sim$240~km/s). In the following, we will adopt for this last quantity the  
velocity of the LSR at the Sun's position.
The location of the same stars in a Toomre diagram, colour-coded with their Galactic orbit  eccentricity is shown in Fig.\ref{Fig:Toomre}. The regions enclosed by the circular dotted lines 
denote those populated by stars with thin disc kinematics, i.e. typical total velocity found within $\pm$40-50~km/s around the LSR value
\cite[see, for instance,][for more details]{PVP_Ale}.
From these figures,
it is clear that most stars have thin disc kinematics as expected, although some \GamDor\ candidates belong actually to the Galactic thick disc or halo (low $V_\phi$-values, high Galactic latitudes, large eccentricities or total velocities). 
Among those having a total velocity larger than $\pm$50~km/s compared with the LSR value (i.e. too large velocities for belonging to the thin disc, first dotted line in Fig.~\ref{Fig:Toomre}), very few have published \gspspec\ atmospheric parameters. Their spectra \SNR\ is indeed low
(around 25) leading to rather large \Vrad\ uncertainty. Moreover,  their rotational rate is high, and their metallicity could be too low for thin disc stars; but, more importantly, the associated \meta\ uncertainties are very large. All of this reveals that their spectra were probably not properly analysed by \gspspec.
In any case, the fact of not belonging kinematically to the thin disc is contradictory to our understanding of the evolutionary stage of this class of variable stars and they will be rejected from the studied sample (see below).
On the contrary, most of the brightest candidates with \gspspec\ parameters (i.e. with the highest \SNR\ spectra) are found closer to the Galactic plane with  kinematics and orbital properties typical of thin disc stars (circular orbits, $V_\phi \sim 215-260$~km/s and/or abs($V_{\rm Tot}-V_{\rm LSR})\la$25~km/s. 

\subsection{\GamDor\ candidates belonging to the thin disc}
\label{Sec:ThinD}
\GamDor\ stars are known to be late A- to early F-type stars, located on the main-sequence below the 
classical
instability 
strip. 
As already mentioned, they
have masses between $\sim$1.3 and $\sim$1.9~M$_\odot$ \citep[see, for instance,][]{Mombarg19, Rhita19,Fritzewski24} and have therefore rather young ages. We refer to
\citet[][see their Fig.\,17]{Mombarg21} who deduced the ages of the 37 best characterised \GamDor\  stars from asteroseismic modelling of their identified oscillation modes and found ages ranging from $\sim$0.15 up to 2\,Gyr.
Similarly, \cite{Fritzewski24} also report asteroseismic ages for 490 \GamDor\ that are always smaller than 3.0~Gyr, their mean age being close to 1.5~Gyr with a dispersion of 0.5~Gyr. 
On the contrary, thick disc stars are older than $\sim$8~Gyr \citep[see, for instance,][]{Michael17, Pablo21, Rix22} whereas halo stars are even older. Therefore, most of the Galactic \GamDor\ stars are expected to belong to the thin disc of the Milky Way. 

One can thus use the above described kinematic and orbital properties of the candidates from \citet{Conny23} to select those with the highest probability of being bona-fide \GamDor\ pulsators. 
This is an entirely complementary and independent selection to the one based on the DR3 or TESS light curves.
We have therefore selected all the above candidates having a high probability to belong to the thin disc, i.e. those having close to circular orbits (eccentricity lower than 0.2)
and ($V_{\rm Tot}-V_{\rm LSR})<$25~km/s. These kinematic criteria\footnote{Adding a filter based on the distance from the Galactic Plane did not modify this selection.} led to the selection of 
2,245 \GamDor\ candidates belonging to the Galactic thin disc (i.e. 83\% of the stars with high-quality Galactic parameters), 385 of them having \gspspec\ parameters. This sub-sample, called the {\it Thin Disc}-sample hereafter, is discussed below. We can therefore conclude that most stars of the initial sample are thin disc members.

\section{Physical parameters of the \GamDor\ pulsators}
\label{Sec:Param} 

All the 650 \GamDor\ pulsators with parameters from \gspspec\ have a published effective temperature (\T) and  a surface gravity (\g). In addition, one can also access their global metallicity (\meta) and abundances in \alfa-elements with respect to iron (\AF) for 602 and 595 of them, respectively. We remind that, within \gspspec, \meta\ is estimated from all the available atomic lines in the RVS spectra and is a good proxy of \FeH. Similarly, \alfa-element abundances are derived from all the available lines of any $\alpha$-elements. The \AF\ abundance ratio is, however, strongly dominated by the huge \CaII\ infrared lines that are present in the \Gaia/RVS wavelength domain and is thus strongly correlated to \CaFe.
Moreover, because of the complex analysis of these rather hot and, usually, 
fast-rotating stars, very few other atomic lines are present in their RVS spectra. Therefore,  individual chemical abundances were derived for only very few tens of them. Only 20 stars have an estimate of their \CaFe\ and the number of stars with other published chemical abundances is even lower. As a consequence, we will only consider hereafter the \AF\ abundance ratio for this \GamDor\ sample.
We calibrated all the above mentioned atmospheric parameters and abundances as a function of \T, adopting the prescriptions recommended in \cite{GSPspecDR3, Ale24}.
In addition, we remind that, associated to the \gspspec\ parameters, there are several quality flags (\flags) that have to be considered to assess the quality of this parametrisation. 

We also adopted other parameters derived from the analysis of the RVS spectra. For example, 
we partially used the \Gaia/DR3 spectral type provided by the $spectraltype\_esphs$ parameters. In addition,
409 of the \gspspec\ \GamDor\ candidates have a published line-broadening measurement \citep[$vbroad$, related to their rotational velocity, see][]{Yves23}, confirming the fast-rotating nature for most of them. We refer to \cite{Conny23} for a detailed study of the \Gaia\ $vbroad$ properties of the $g$-mode pulsators.
Even if $vbroad$ was in the end not published for most stars in \Gaia/DR3,
\gspspec\ has published three quality flags \cite[$vbroadT, vbroadG, vbroadM$, see Tab.~2 in][]{GSPspecDR3} that depend directly on $vbroad$ (internally delivered within DPAC for the spectra analysis). The possible biases in the \gspspec\ parameters that could be induced by rotational line-broadening can therefore be explored by future users thanks to these three $vbroadTGM$ flags. 
Contrarily to $vbroad$, 
these flags are available for the whole sample and are used below to complement the rotational broadening information, when necessary. In the following and for convenience, all these line-broadening quantities will be referred to as {\it rotational velocities}.

Finally, we remind that the \Gaia/DR3 spectroscopic  parameters were derived by the \gspspec\ module by assuming that the rotation rate of the analysed stars are rather low. Therefore, highly rotating stars were rejected: depending on the
stellar type, $vbroad$ limitations are around $\sim$30-40~km.s$^{-1}$, and the parametrisation quality degrades quickly above $\sim$25~km.s$^{-1}$.
Because of this parametrisation limitation, we warn the reader that the \GamDor\ candidates parametrised by \gspspec\ have thus lower rotational velocities than typical values for these variable stars. 
Indeed, the $vbroad$ distribution of our stars has a mean value around $\sim$25~km.s$^{-1}$, associated to a standard deviation equal to $\sim$10~km.s$^{-1}$, and a maximum value of 58~km.s$^{-1}$. As a comparison, 
these stars are known to rotate at 40-100~km.s$^{-1}$ and some can reach up to $\sim$150~km.s$^{-1}$, see for instance \cite{Gebruers21, Conny23}. This rotational velocity of the stellar atmosphere is related to the internal rotation rate, as discussed for instance by \cite{GangLi20}. As a consequence, we are conscious that our sample is biased towards \GamDor\ with rather low rotation rate, and should therefore not be fully representative of these specific class of variable stars. \\

{\it Stellar luminosities and radii:}
To complement this spectral parametrisation, we computed the luminosity ($L_\star$) and radius ($R_\star$)
for each star.
For that purpose, we first estimated the extinction $E(B_P-R_p)$ in the \Gaia\ bands by subtracting the observed $(B_P-R_p)$ colour from a {\it theoretical} one. The latter was calculated from the \gspspec\ \T, \g\ and \meta, inverting the \cite{Luca21} relation that predicts stellar colour from atmospheric parameters \citep[see also Sect.~10.2 in][]{GSPspecDR3}. This procedure did not converge for a few stars, hence these were rejected hereafter.
We then estimated the coefficients $k_{TGMA} =  A_G / E(B_P-R_p)$, $A_G$ being the absorption in the $G$-band. These coefficients depend on the four atmospheric parameters and have been estimated thanks to the tables provided with the \Gaia\ stellar parameters\footnote{https://www.cosmos.esa.int/web/gaia/dr3-astrophysical-parameter-inference}. The value of $A_G$ then allows to deduce the absolute magnitude in the $G$-band
from the \Gaia\ DR3 $G$-magnitude and the \cite{Coryn21} geometric distances.
We derived the luminosities, adopting the bolometric corrections ($BC$) from \cite{Luca18}. We note that the considered relations to estimate $k_{TGMA}$
and $BC$ do not depend on \AF. When available, we adopted the relation of \cite{Salaris93} to include the $\alpha$-element content into the global metallicity.
Finally, thanks to the \gspspec\ \T\, one directly obtains the stellar radius. The quality of these radii simply computed from \Gaia\ photometry, distances and spectroscopic parameters is excellent.
We refer to de Laverny et al. (2024, to be submitted) and \cite{Ale24} for a detailed comparison with interferometric and/or asteroseismic radii, that confirmed the high-quality of our $R_\star$ estimates.
One could then also get the stellar mass ($M_\star$) from the surface gravities, but we favoured to fix this quantity thanks to the known typical masses of \GamDor\ (see below the discussion on the adopted \g). For all these parameters, the uncertainties were estimated by performing 1000 Monte-Carlo
realisations, propagating the uncertainties on each atmospheric parameters (that reflect the \SNR\ spectra), distance
and \Gaia\ magnitudes. \\

{\it Effective temperatures:}
Since \GamDor\ stars are known to be early-F spectral type and in order to define a sub-sample of 
high-quality parametrised stars, we first checked their spectral type provided by the \Gaia/DR3.
Among the 650 candidates parametrised by \gspspec, 562, 82 and 6 were found by the DPAC/ESP-HS module to belong to the 'F', 'A' or 'B' spectral types, respectively. The \gspspec\ \T\ also confirmed the too hot temperature with respect to typical \GamDor\ stars of a few other stars \citep[see, for instance,][]{VanReeth15,Conny23}, hence they were filtered out. 
Moreover, we also found that $\sim$6\% of the candidates have a \gspspec\ \T\ much cooler and not compatible with an early spectral type. All the spectra of these outliers suffer from large \Vrad\ uncertainties and/or very large rotational velocities and/or low \SNR\ that may lead to an erroneous parametrisation. Therefore, all these too cool stars were also rejected. The remaining 598 spectroscopic candidates have a \Gaia\ $(B_p-R_p)_0$ colour (corrected from extinction by us) fully compatible with
their \T. The median of their $(B_p-R_p)_0$ colour is 0.5 and the associated dispersion is found to be extremely small (0.04~mag.), confirming their early F-type nature.
These 598 stars will be called {\it F-type} \gspspec\ sample, hereafter.\\

{\it Stellar surface gravities:} Knowing the stellar luminosity, the \gspspec\ effective temperature and adopting a \GamDor\ typical mass,
the surface gravity can simply be estimated for almost all the \gspspec\ pulsators (called \g$_{\rm Lum}$, hereafter). Practically, we randomly chose the mass of each star within the \GamDor\ mass range 1.3 to 1.9~$M_\odot$ \citep[see, for instance,][]{Fritzewski24}, assuming an uniform distribution. The associated mass uncertainty has been fixed to half of this range ($\pm$0.3~$M_\odot$).
We note that varying the mass over the entire range covered by genuine \GamDor\ stars changes our  \g$_{\rm Lum}$ estimates by about 0.1~dex and, thus, do not affect our conclusions. The \g$_{\rm Lum}$ uncertainties were computed from 1000 Monte-Carlo realisations, propagating the uncertainties on $L_\star$, \T\ and $M_\star$.
 We finally point out that 30 stars (among the 650) have no \g$_{\rm Lum}$ because no extinction was available for them (see above). 
 
 It was found that most of these \g$_{\rm Lum}$ are in very good agreement with the \gspspec\ \g\ for stars with a rather low rotational velocity ($vbroad \la$15-20~km/s) or for stars with slightly larger rotational rate but with high \SNR\ spectra ($\ga$100). 
However, larger discrepancies between the two surface gravity estimates are found for the highest rotators and/or stars with low \SNR\ spectra (mean difference of log($g_{\rm \gspspec}/g_{\rm Lum}$)$\sim$0.35~dex, with a standard deviation of 0.4~dex). This results from the DR3 \gspspec\ pipeline that is optimised for non-rotating stars and some parameter biases could exist for stars with large rotation rates\footnote{This will be addressed for \Gaia/DR4.}

In the following and in order to avoid surface gravities potentially affected by the stellar rotation,
we adopted these \g$_{\rm Lum}$ and their associated uncertainties. Thanks to this procedure, we have a sub-sample of about 600 \GamDor\ pulsators with rather accurate surface gravity (found in the range from $\sim$3.5 to $\sim$4.5) and effective temperatures typical of \GamDor\ stars, according to \cite{VanReeth15, Fritzewski24}. We emphasize that all the above derived quantities do not rely on any stellar isochrones, but only on \Gaia\ astrometric and photometric data and RVS spectra.\\

\begin{figure*}[t]
    \centering
    \includegraphics[width=0.49\textwidth]{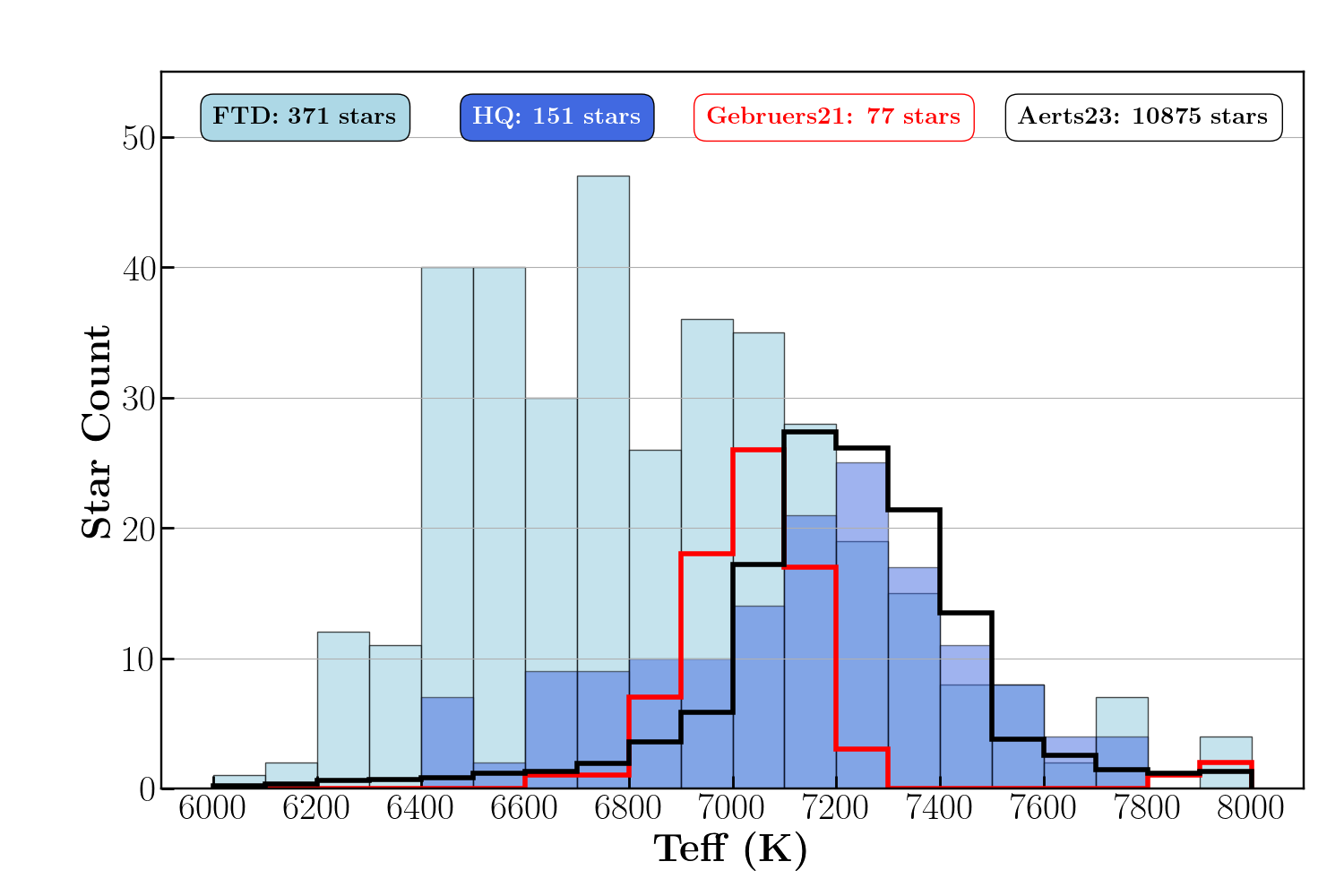}
    \includegraphics[width=0.49\textwidth]{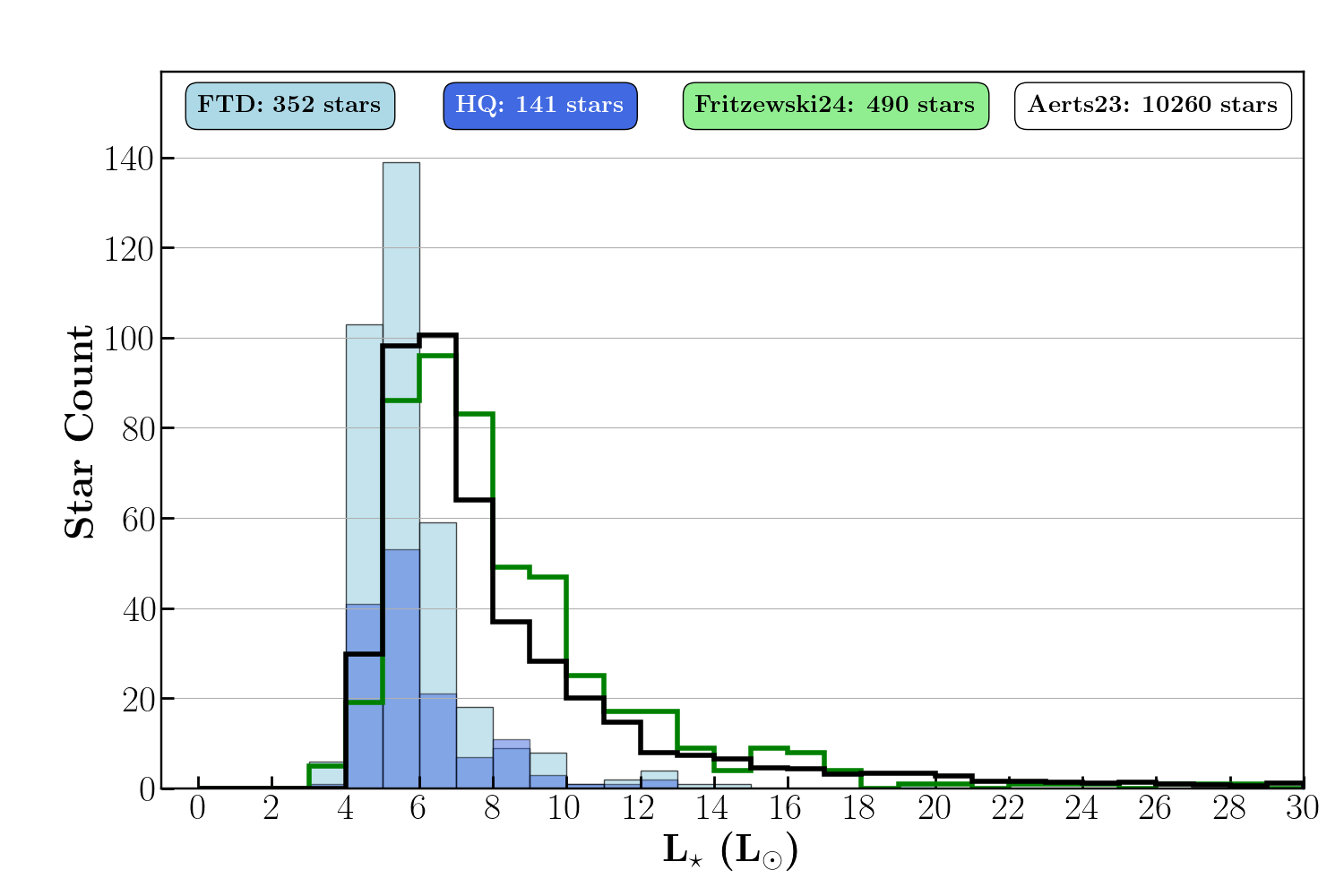}    
    \includegraphics[width=0.49\textwidth]{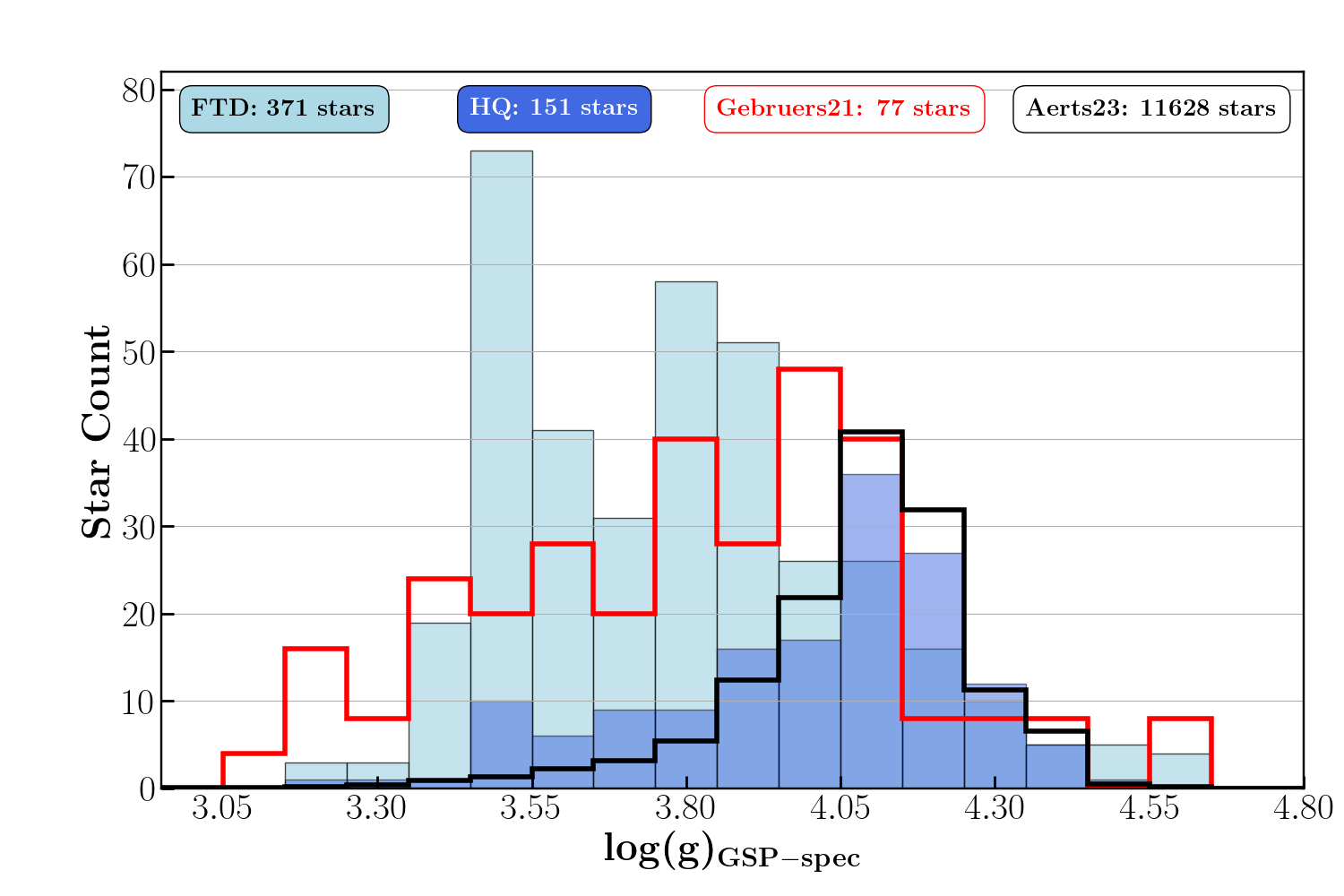}
    \includegraphics[width=0.49\textwidth]{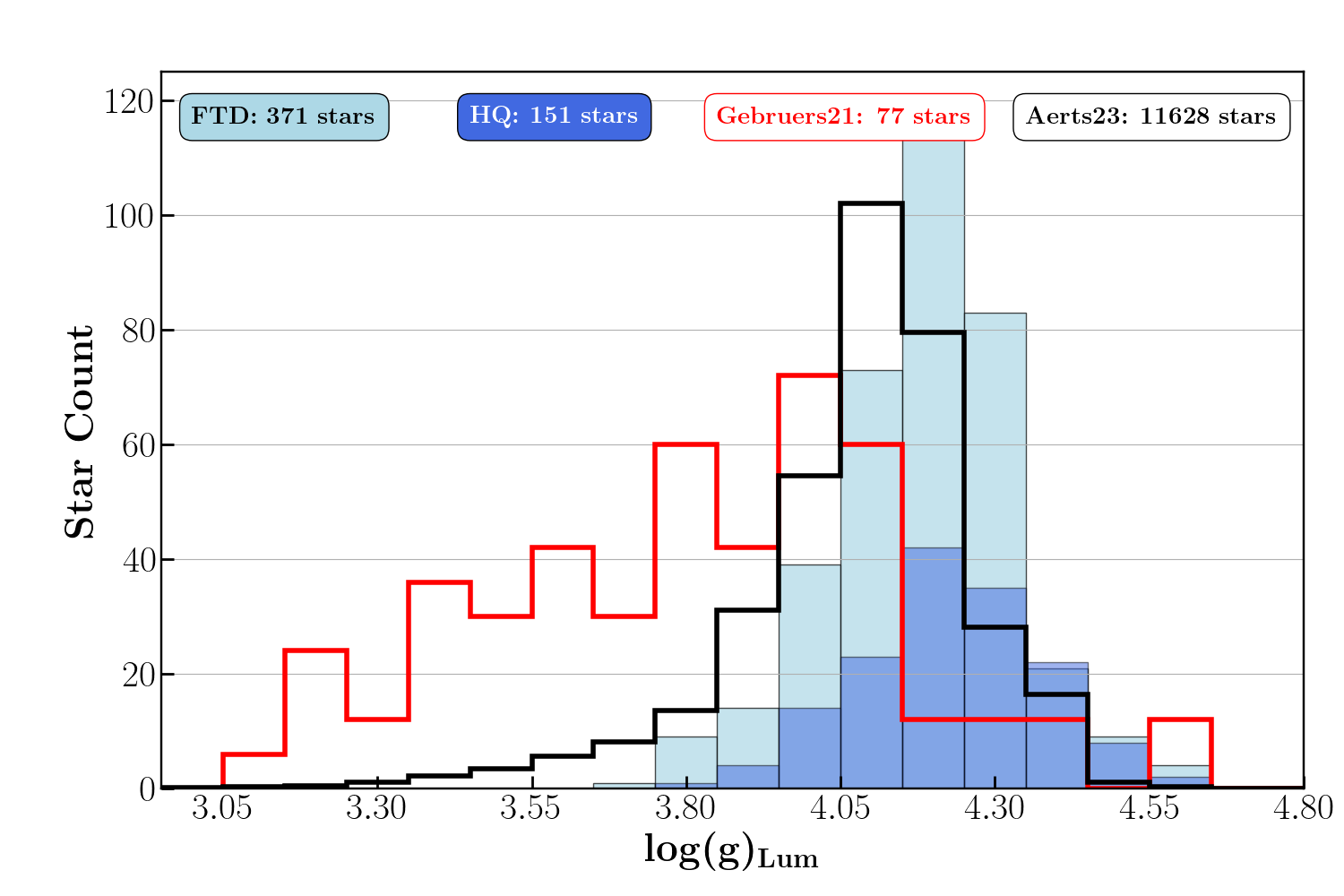}
    \includegraphics[width=0.49\textwidth]{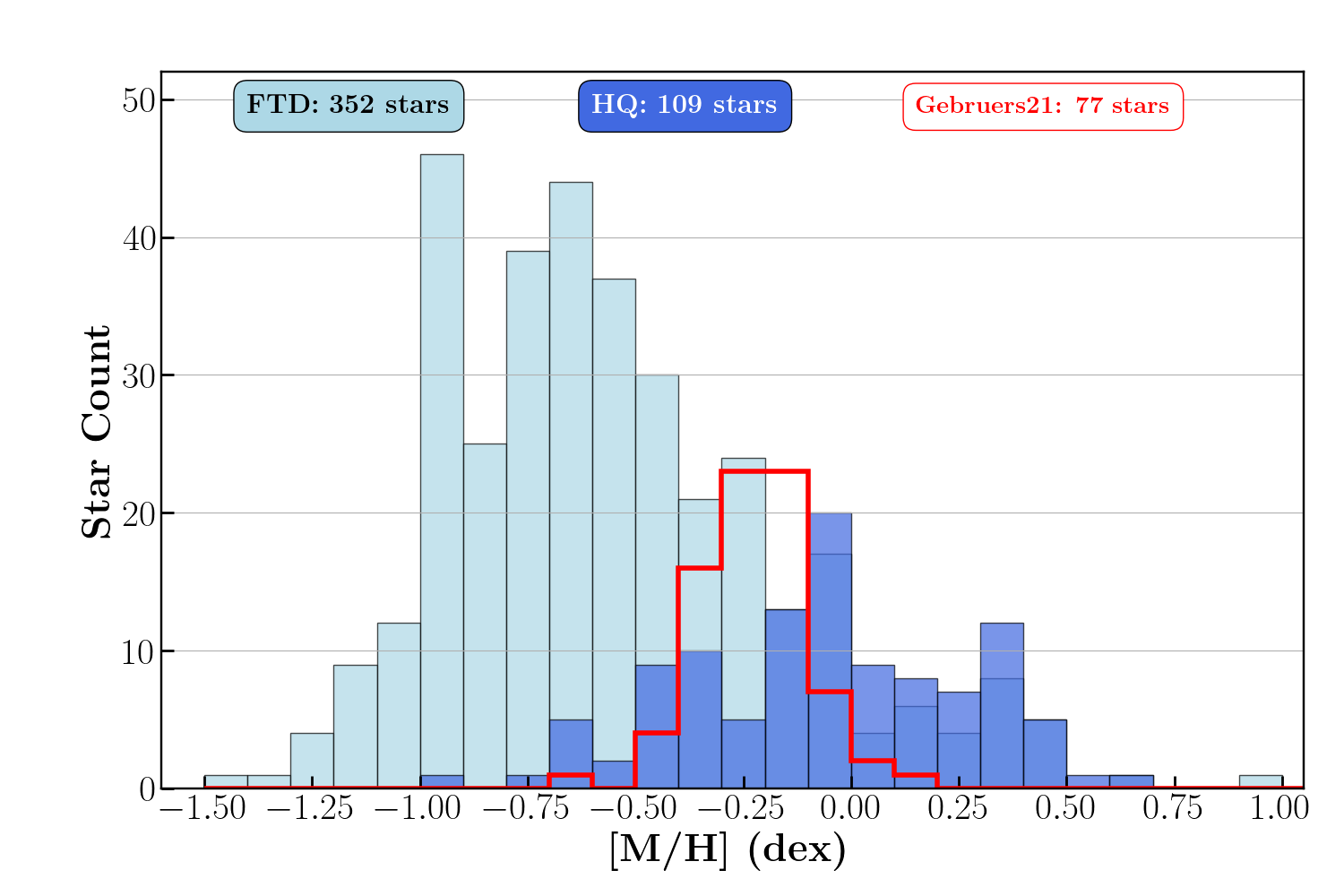}
    \includegraphics[width=0.49\textwidth]{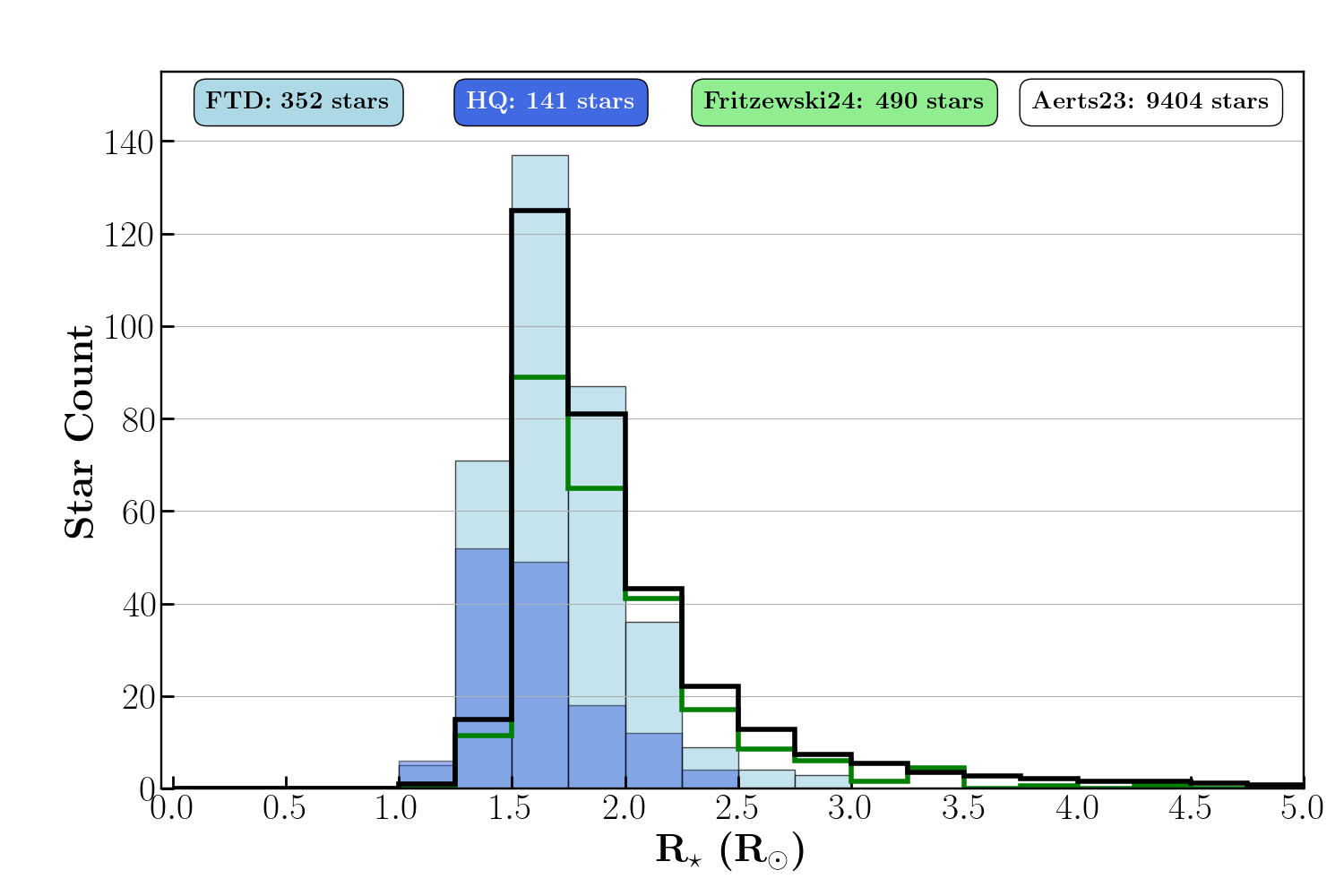}
    \caption{Distribution of the \gspspec\ \GamDor\ effective temperatures, luminosities, surface gravities, global metallicities and stellar radii for the {\it F-type Thin Disc} ($FTD$, light-blue histograms) and {\it High-Quality} ($HQ$, in dark-blue) sub-samples (see text for details). The red, black and green distributions are for the parameters derived by \cite{Gebruers21}, \cite{Conny23} and \cite{Fritzewski24}, respectively,     
    artificially scaled since their statistics differ from ours. \g$_{\rm GSP-spec}$ refers to the stellar surface gravities derived from the spectra analysis by \gspspec\ whereas \g$_{\rm Lum}$ is the estimation from the stellar luminosity, assuming typical \GamDor\ masses. The number of stars in each panel may differ since not all the parameters are available for every sources.}
    \label{Fig:Histo}
\end{figure*}

{\it Global metallicities:} We simply adopted the \gspspec\ calibrated metallicities for all the \GamDor\ pulsators having a \T\ and a \g$_{\rm Lum}$, as defined above.\\

{\it Parameter uncertainties:} The RVS spectra of all these \GamDor\ pulsators belonging to the
 {\it F-type} \gspspec\ sample
have rather low \SNR\ ratios. The median of these \SNR\ is 34 and only 9\% of the spectra have a \SNR>100. 
The resulting median uncertainties\footnote{Estimated by propagating the RVS spectra flux errors on the parameter determination through Monte-Carlo realisations \citep[see][]{GSPspecDR3}.} on (\T, \g, \meta, $L_\star$, $R_\star$) are therefore rather high (182~K, 0.13~dex, 0.21~dex, 0.13~L$_\odot$, 0.11~R$_\odot$) with a dispersion although still reasonable of (62~K, 0.02~dex, 0.07~dex, 0.07~L$_\odot$, 0.04~R$_\odot$).

\section{Properties of the \Gaia\ \GamDor\ pulsators}
\label{Sec:properties}
The distribution of the main stellar parameters derived above \T, \g\ (both derived from the spectrum analysis and from the luminosity), \meta, $L_\star$, and $R_\star$ are shown in Fig.~\ref{Fig:Histo}. 
In the different panels of this figure, we considered the sub-sample of 371 slowly-rotating stars belonging to both the {\it F-type} \gspspec\ and {\it Thin Disc} samples, that is those having effective temperature typical of \GamDor\ stars and having typical thin disc kinematics as defined in Sect.~\ref{Sec:ThinD}. This sub-sample will be called {\it F-type Thin Disc} ($FTD$) hereafter. It is shown as light-blue histograms in Fig.~\ref{Fig:Histo} and should contain good \Gaia\ \GamDor\ candidate stars.

For comparison purposes, we have over-plotted in Fig.~\ref{Fig:Histo} the parameter distributions derived by three recent studies that provide parameters for large numbers of \GamDor\ stars. First, \citet[red-line distributions in Fig.~\ref{Fig:Histo}]{Gebruers21} report \T, \g\ and \meta\ 
estimated from the analysis of high-resolution spectra ($R \sim$ 85,000)
for 77 bona-fide \GamDor\ stars with asteroseismic modelling from {\it Kepler\/} observations. None of these stars were found in the  {\it F-type} \gspspec\ sample. Secondly, \cite{Fritzewski24} published luminosities and asteroseismic radii of 490 \GamDor\ derived from \Gaia\ and {\it Kepler\/} observations. 
Their parameter distributions are shown as green lines in Fig.~\ref{Fig:Histo}. There is only one star of \cite{Fritzewski24} that was parametrised by \gspspec, but it is not included in the  {\it F-type} \gspspec\ sample because of its too large rotational broadening and associated low-quality flags.
Finally, we also show the \T, \g, $L_\star$ and $R_\star$ distributions of the \GamDor\ candidates of \cite[black histograms]{Conny23} as another comparison. Some of their stars suffer from rather large uncertainties on their parameters and we therefore filtered out the stars with \T\ and \g\ errors larger than 100~K and 0.2~dex, respectively, before constructing the black histograms. Moreover,  stars with relative uncertainties in $L_\star$ and $R_\star$ larger than 25\% and 50\% were also rejected. \\

\begin{figure}[t]
        \centering
        \includegraphics[width=0.49\textwidth]{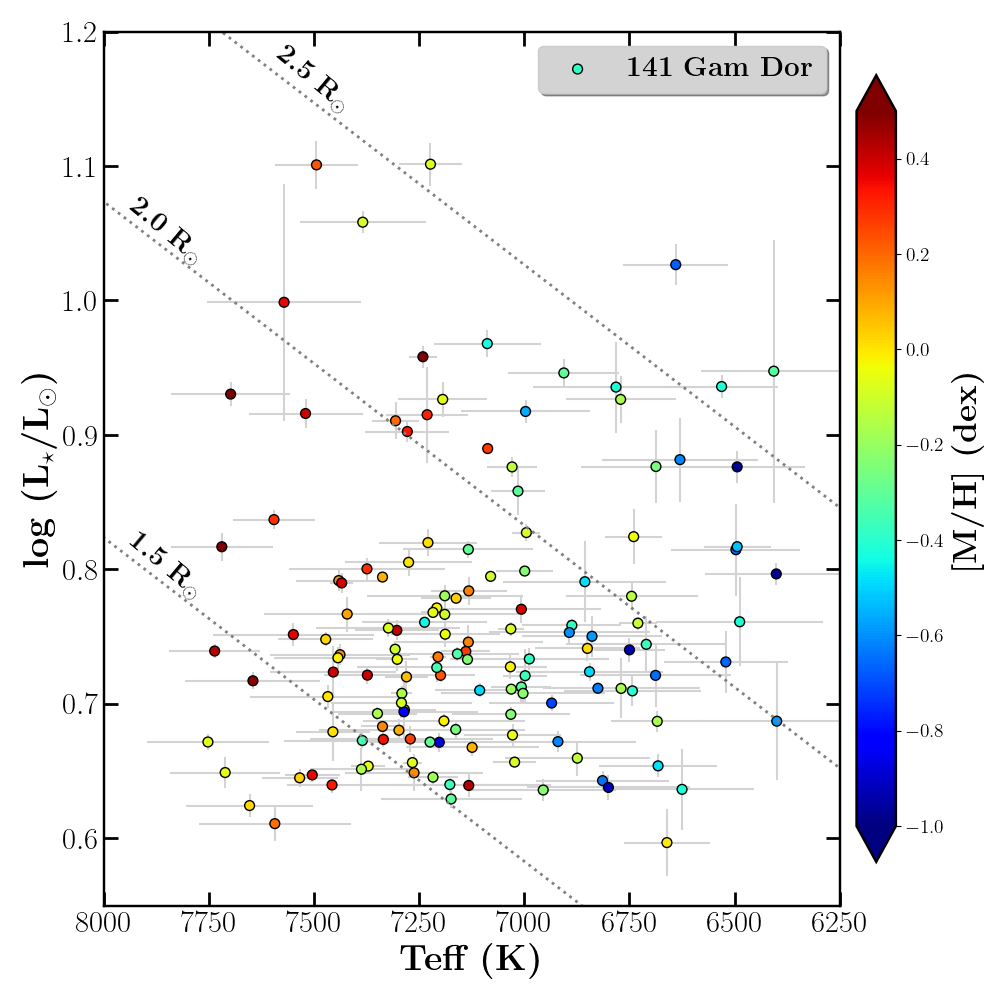}  
        \caption{Luminosity versus effective temperature diagram of the $HQ$ sub-sample of \GamDor\ stars, colour-coded with their global metallicity. Error bars on $L_\star$ and \T\ are shown in light-grey and the dotted lines represent the iso-radius relations.}
        \label{Fig:LumTeff}    
\end{figure}

Some other \GamDor\ stars were also analysed by high-resolution spectroscopy as, for instance, in \cite{Tkachenko12, Niemczura15, Niemczura17}. The spectra collected by these authors have a resolution of 32 000, 85 000, and 25 000/45 000, respectively, but their sample are much smaller than ours. None of our stars are found in these samples but one discusses below the atmospheric parameter properties derived by these studies with respect to the \gspspec\ ones (see also the discussion about Fig.~\ref{Fig:LumTeff}).\\





Regarding the effective temperatures, it can be seen in Fig.~\ref{Fig:Histo} that 
our \T\ distribution is nearly compatible with the estimates reported in the literature.
We do report a large number of cooler \GamDor\ stars than \cite{Gebruers21} and \cite{Conny23}. We checked that  most of these cooler stars had bad quality \gspspec\ $vbroadTGM$ flags,
meaning that their parametrisation was not optimal because of the broadened lines in their spectra. Moreover, we remind that most RVS \GamDor\ spectra are of rather low-quality, leading to large \T\ uncertainties. As a consequence, if one selects only
the 151 stars belonging to the {\it F-type} \gspspec\ sample and having a \T\ error less than 200~K (the mean \SNR\ ratio of the selected spectra is then around 60) plus their three $vbroadTGM$ flags strictly smaller than 2, we then obtain a \T\ distribution fully compatible with the two comparison ones. This {\it High-Quality} ($HQ$) sub-sample of \GamDor\ stars is shown as dark-blue histograms in Fig.~\ref{Fig:Histo}. Our \T\ distribution is also fully compatible with the ones derived from high-resolution spectroscopy (see above cited references), if one excludes binaries and/or hybrid stars: their \T\ are always found in the $\sim$6700 -- $\sim$8000~K range. \\

Our stellar luminosity distributions ($FTD$ and $HQ$ sub-samples) are peaked around 5~L$_\odot$ with few bright stars up to $\sim$25~L$_\odot$. They are close to the distribution of the two comparison samples (similar peak in $L_\star$) although \cite{Conny23} and \cite{Fritzewski24} report a larger number of more luminous stars. We have checked that considering only stars with good \Gaia\ astrometric $ruwe$ parameters ($ruwe$ < 1.4) in the different comparison samples
(about 10\% stars would be rejected) does not modify these distributions. Moreover, we remind that these two other studies computed their stellar luminosity adopting a \Gaia/DR3 interstellar reddening 
that could differ from our own absorption estimate. Since we did not find any
specific differences between these two reddening flavours, the lack of high-luminosity stars in our sample could be due to selection bias effects: either a lack of too hot and, hence more luminous \GamDor\ stars, not parametrised by \gspspec; and/or a lack of low surface gravity stars, hence with high-radius and luminosity, as seen in the \g\ and radius panels of Fig.~\ref{Fig:LumTeff}. \\


The \gspspec\ $FTD$ and \cite{Gebruers21} spectroscopic surface gravities are
in good agreement (left panel, second row in Fig.\ref{Fig:Histo}). \GamDor\ surface gravities derived 
by other analysis of high-resolution spectra studies also agree, although we note that the samples of \cite{Niemczura15, Niemczura17} have \g\ within $\sim$3.6--$\sim$4.0, i.e. surface gravities slightly smaller than ours and those of \cite{Gebruers21} that reach up to \g=4.4. However, more numerous lower gravity values are seen in Fig.~\ref{Fig:Histo} compared to \cite{Conny23}. Such stars with a surface gravity lower than $\sim$3.5 should have left the main sequence (or are close to this evolutionary stage).
The agreement between the $HQ$ spectroscopic gravities and those of \cite{Conny23} is however excellent, confirming again that low-quality and line-broadened spectra are more difficult to parametrise.
Moreover, the results from \cite{Conny23} are also in good agreement with our surface gravities estimated from the stellar luminosity (\g$_{\rm Lum}$ shown in the right panel, second row) and assuming \GamDor\ masses in the range 1.3-1.9~$M_\odot$. This is quite normal since the \cite{Conny23} \g\ values were estimated by the DPAC GSP-phot module \citep{Rene23} from the luminosity, \T\ and stellar isochrones, a rather similar method to ours, except for using isochrones. From these different comparisons, we conclude that \GamDor\ surface gravities derived from the stellar luminosity and assuming typical \GamDor\ masses should be preferred over the spectroscopic ones because of the difficulty to analyse spectra of fast-rotating stars, except if one considers the $HQ$ stars for which the spectroscopic gravities are excellent. \\

As for the \GamDor\ stars' mean metallicities, our $FTD$ sample covers a larger range in \meta\ than the comparison samples. More importantly, it reveals an excess of low-\meta\
stars  with respect to the spectroscopic sample of \cite{Gebruers21}. The latter peaks around -0.2~dex, whereas ours peaks at about -0.7~dex.
We note that the confirmed \GamDor\ of \cite{Tkachenko12} with parameters derived from high-resolution spectra have \meta\ that cover a distribution close to the one of \cite{Gebruers21}.
We remind that such low-metallicities are not expected to be numerous for thin disc stars. This shift towards lower metallicities could again be partially explained by the stellar rotation and/or low-quality RVS spectra, as it might induce a bias in the parametrisation. 
Again, rejecting such complex spectra and considering $HQ$ sub-sample stars with a  \meta\ uncertainty less than 0.25~dex leads to a metallicity distribution that is more compatible with \cite{Gebruers21}.
We still have however more stars with higher and lower metallicities than them by about $\pm$0.5dex. They could be present since our sample has a larger spatial coverage in the Milky Way.\\

Finally, the compared stellar radius $FTD$ and $HQ$ distributions are in very good agreement with the asteroseismic radii determined by \cite{Fritzewski24}, which are expected to be the most accurate ones. 
The spectroscopic radii, whatever the sample considered, are indeed very well representative of those expected for \GamDor\ stars.
Our radius median value is close to 1.7~$R_\odot$ and 90\% of the sample is found in the range
1.35 -- 2.35~$R_\odot$. The largest stars in our sample have $R\sim3.5-4~R_\odot$, so they must be rather close to leave the main-sequence, which is confirmed by their 
effective temperatures and high luminosities (\T$\simeq$7000~K and $L \simeq$25~$L_\odot$).\\

\begin{figure}[t]
        \centering
        \includegraphics[width=0.49\textwidth]{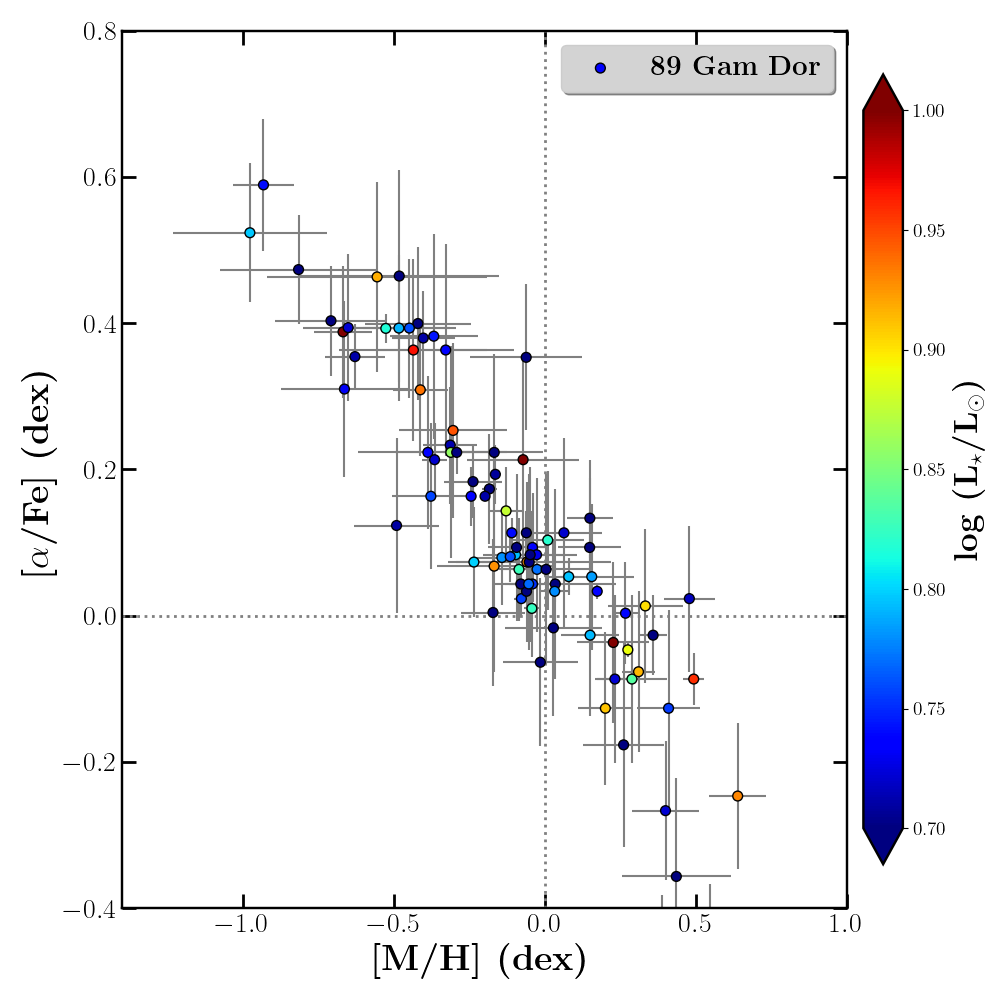}   
        \caption{Distribution of the $HQ$ sub-sample of \GamDor\ stars in the \AF\ versus \meta\ plane. Only stars with an uncertainty smaller than 0.15~dex in \AF\ are shown. The Solar location is indicated by the intersection of the dotted lines.}
        \label{Fig:Alpha}    
\end{figure}

In summary, our reported \T, \g, \meta, $L_\star$ and $R_\star$ are in very good agreement with recent literature values, in particular when considering the {\it High-Quality} sub-sample defined by selecting high-quality spectra or low-rotating stars. Furthermore, our $L_\star$, 
\g$_{\rm Lum}$, and $R_\star$ values are trustworthy, even without considering the membership of the Galactic thin disc as an extra good measure of the star being a genuine \GamDor\ pulsator and/or their spectra properties. 

Moreover, since we have shown that the agreement on the effective temperature and mean metallicity is improved by
selecting high-quality \SNR\ stellar spectra or stars without large rotational broadening velocities, 
we show these $HQ$ \GamDor\ stars 
in a luminosity - effective temperature  diagram (Fig.~\ref{Fig:LumTeff}), 
colour-coded with their metallicity.
This figure is similar to those found in the literature, as for example in \cite{Niemczura15, Niemczura17, Tkachenko12, Gebruers21}. Excluding binaries and hybrid stars, all our \GamDor\ stars are very well concentrated in the same small region of the $L$-\T\ diagram (or \g-\T, in some of the above cited works).
We note that there is no \GamDor\ stars in this figure with \T\  hotter than $\sim$7750~K\footnote{A few hotter \GamDor\ can be found in the literature \citep[see, for instance,][]{Kahraman20bis}}. This bias is caused by the \gspspec\ parametrisation that was optimised for 
FGKM-type stars (we recall that the reference grid 
is based on spectra models cooler than 8000~K).
We therefore cannot exclude that hotter \GamDor\ could exist but they were rejected during the parameterisation (this will be updated for \Gaia/DR4).
Finally, it can be seen in Fig.\ref{Fig:LumTeff} that more metal-rich  \GamDor\ are found at higher \T, whatever their luminosity is. This could be partly due to some possible parametrisation biases: for instance, metal-poor hot star spectra show very few lines and are thus more difficult to parameterise particularly when their rotation rate is high, explaining probably the absence of such stars in the present sample. But this could also be real and could be a signature of the different evolution of stars with slightly different masses and metallicity. For instance, by exploring BaSTI evolutionary tracks \citep{BaSTI}, it can be seen that metal-rich \GamDor\ stars with masses around 1.6-1.9~$M_\odot$ appear hotter than lower mass ($\sim$1.3~$M_\odot$) more metal-poor stars. More specifically, we have estimated that a difference of $\sim$0.3~$M_\odot$ implies a shift in \T\ of similar amplitude as a difference of $\sim$0.7~dex in metallicity. This rather well corresponds to what is seen in Fig.\ref{Fig:LumTeff}. 

The same $HQ$ sub-sample but filtering out stars with \AF\ uncertainty greater than 0.15~dex is plotted in an \AF\ versus \meta\ diagram in Fig.~\ref{Fig:Alpha}. Such  a filtering again reduces the number of stars
but reveals more accurate chemo-physical properties of \GamDor\ pulsators.
We remind that only the global \AF\ abundances (which is a good indicator of [Ca/Fe] for RVS spectra) of these \GamDor\ stars parametrised by \Gaia\ \gspspec\ are available in the DR3 catalogue because of too low statistics of individual chemical abundances (see above). Fig.~\ref{Fig:Alpha}
confirms that the selected sample has chemical properties consistent with the Galactic disc population, that is a constant decrease of \AF\ with the metallicity for \meta>-1.0~dex. Moreover, since the membership
of most of these stars to the thin disc was based on purely kinematics and dynamical criteria, such a \AF\ versus \meta\ trend is an independent proof that the chemo-physical  properties of the \GamDor\ pulsators can be safely adopted.

Thus, the \gspspec/DR3 analysis of \GamDor\ pulsators provides useful physical and/or chemical parameters once an optimized filtering of the poorly parametrised spectra (low \SNR\ or too fast rotating stars) is performed. The parameters of these $HQ$ \GamDor\ stars are provided in an electronic table which content is presented in Table.~\ref{Tab:HQ}.

\begin{table}[t]
        \caption{\label{Tab:HQ} \gspspec\ parameters of the {\it High-Quality} \GamDor\ stars.}
        \centering
        \begin{tabular}{ll}
        \hline
        Label & Description \\
        \hline
        GDR3id & \Gaia\ DR3 source ID\\
        Teff & Effective temperature (K)\\
        Teff\_err &  \T\ uncertainty (K) \\        
        logg &  Stellar surface gravity ($g$ in cm.s$^{-2}$) \\
        logg\_err & \g\ uncertainty (dex) \\ 
        Meta & Mean metallicity (\meta$\sim$[Fe/H],  in dex)\\
        Meta\_err & \meta\ uncertainty (dex) \\
        AFe & Enrichment in \AF\ (dex)\\
        AFe\_err & \AF\ uncertainty (dex) \\
        L & Stellar luminosity ($L_\odot$)\\
        L\_err &  $L_\star$ uncertainty ($L_\odot$) \\
        R &  Stellar radius ($R_\odot$) \\
        R\_err &  $R_\star$  uncertainty ($R_\odot$)  \\
        TD &  Thin disc membership (Yes=1)\\
         \hline
        \end{tabular}
\end{table}



\section{Conclusions}
\label{Conclu}

We have studied the \Gaia/DR3 spectroscopic parameters derived from the analysis of the RVS spectra for the large sample of \GamDor\ candidate pulsators composed by \cite{Conny23} and confirmed in \citet{HeyAerts24}. About 38\% of these stars have a published radial velocity and $\sim$6\% of them were actually analysed by the \gspspec\ module in charge to analyse their \Gaia\ spectra.

Thanks to the available \Vrad\ and astrometric \Gaia\ information, we have been able to compute kinematics and orbital information for all these stars. This allowed us to identify that  2,245 of them (i.e. most of the candidates with high-quality kinematics) belong to the thin disc of the Milky Way, which is expected since these gravity mode pulsator should have typical ages lower than 2-3~Gyr.

We then computed their luminosity and stellar radius from \Gaia\ astrometric and photometric data, adopting the \gspspec\ effective temperature
and without considering any stellar evolutionary models or isochrone priors.
A comparison with recently published values of well studied \GamDor\ stars reveals
that the derived luminosities, stellar surface gravities derived from $L_\star$ and
assuming typical \GamDor\ masses, as well as the stellar radii, are of high quality.
Moreover, a strict filtering rejecting stars with large \T\ uncertainty 
(caused by too low \SNR\ RVS spectra) or high rotational velocity led to pulsators 
with the best derived parameters, including \T, \meta, and \AF . All of these observables were found to be fully consistent with typical values of genuine slowly-rotating \GamDor\ pulsators. Indeed, the \gspspec\ {\it High-Quality} \GamDor\ stars have effective temperatures between $\sim$6,500 and  
$\sim$7,800~K and surface gravities around 4.2.
Their luminosities and stellar radii peak at $\sim$5~L$_\odot$ and $\sim$1.7~L$_\odot$, whereas their metallicity distribution is centered close to the Solar value, covering the range [-0.5, +0.5]~dex. 
Their \AF\ properties 
 is consistent with the chemical properties of the Galactic disc population. We note that the final number
of parametrised stars is smaller compared to the initial sample because of the low \SNR\ spectra of many of  them, together with the fact that most of them 
are fast rotator for which the \gspspec\ analysis pipeline was not optimised for the \Gaia/DR3. Anyway, the number of newly spectroscopically parametrised \GamDor\ presented in this work is about a factor two larger than in previous studies. \\


Finally, it can be concluded that the \gspspec\ analysis of \GamDor\ stars provides a significant added value to the study of these gravity-mode pulsators, delivering their physical and chemical properties.
This will be even more important with the future \Gaia\ data releases for which the RVS spectra \SNR\ values and the number of analysed stars will be significantly increased. 
Indeed, it is expected that the \SNR\ increase between DR3 spectra and those released in DR4 and DR5 will correspond to a factor
$\sqrt2$ and 2, respectively.
Moreover, it is also anticipated that the analysis of fast-rotating stars by the \gspspec\ module will be improved by considering reference grids of synthetic spectra representative of a wide variety of stellar rotational values, contrarily to the present study that is  biased towards slowly-rotating \GamDor. Finally, hot star spectra will also be considered for reference, allowing a better  parametrisation for stars with high \T. These anticipated future improved analyses of gravity-mode pulsators performed by the \gspspec\ module will therefore be 
of prime interest as input for asteroseismology of such stars. They will indeed allow to define a much larger and thus more statistically significant number of bona-fide \GamDor\ stars with physical and chemical properties. Furthermore, they will also allow to study the  spectroscopic parameters of hotter gravity-mode pulsator, such as the Slowly Pulsating B stars with the aim to improve their asteroseismic modelling \citep{Pedersen21}.

\begin{acknowledgements}
This work has made use of data from the European Space Agency (ESA)
mission \Gaia\ (https://www.cosmos.esa.int/gaia), processed by the \Gaia\ Data Processing and Analysis Consortium (DPAC, https://www.cosmos.esa.int/web/gaia/dpac/consortium). Funding for the DPAC has been provided by national institutions, in particular the institutions participating in the \Gaia\ Multilateral Agreement.\\
This work has also made use of the SIMBAD database, operated at CDS, Strasbourg, France  \citep{Simbad}, the IPython package \citep{ipython}, NumPy \citep{NumPy}, Matplotlib \citep{Matplotlib}, Pandas and TOPCAT \citep{Topcat}.
PdL and ARB acknowledge funding from the European
Union’s Horizon 2020 research and innovation program under SPACE-H2020
grant agreement number 101004214: EXPLORE project.
CA acknowledges funding 
from the European Research Council (ERC)
under the Horizon Europe programme (Synergy Grant agreement
number 101071505: 4D-STAR project).  While partially funded by the European
Union, views and opinions expressed are however those of the authors
only and do not necessarily reflect those of the European Union or the
European Research Council. Neither the European Union nor the granting
authority can be held responsible for them.

Finally, we are
grateful to the anonymous referee for their constructive
remarks.
\end{acknowledgements}

\bibliographystyle{aa} 
\bibliography{ref}

\end{document}